\documentclass[reprint,superscriptaddress,amsmath,amssymb,aps,prl]{revtex4-2}
\pdfoutput=1
\usepackage{amsmath}
\usepackage{amsfonts}
\usepackage{amssymb}
\usepackage{amsthm}
\usepackage{bm}
\usepackage{color}
\usepackage{dcolumn}
\usepackage{graphicx}
\usepackage{esint}

\usepackage[dvipsnames]{xcolor}
\usepackage{xcolor}
\usepackage[normalem]{ulem} 

\newcommand{\g}[1]{{\color{ForestGreen}#1}}

\usepackage{mathtools}

\usepackage{hyperref}
\hypersetup{colorlinks=true,linkcolor=blue,filecolor=magenta,urlcolor=cyan}
\usepackage{textgreek}
\usepackage[utf8]{inputenc}

\usepackage{microtype}   

\begin{document}
\title{Osmotically Induced Shape Changes in Membrane Vesicles}

\author{Rajiv G Pereira}
\affiliation{Department of Physics, Kansas State University, 1228 M. L. K. Jr. Drive, Manhattan, KS, 66506, USA.}

\author{Biswaroop Mukherjee}
\affiliation{NewOrbit Space Ltd, 11 S View Park, Marsack St, Reading RG4, UK.}
\affiliation{Department of Physics and Astronomy, University of Sheffield, Sheffield S3 7RH, UK.}

\author{Sanjeev Gautam}
\affiliation{Department of Physics, Kansas State University, 1228 M. L. K. Jr. Drive, Manhattan, KS, 66506, USA.}

\author{Mattiangelo D'Agnese}
\affiliation{Department of Physics and Astronomy, University of Sheffield, Sheffield S3 7RH, UK.}
\affiliation{Kaplan International College, Palace House, 3 Cathedral Street, London SE1 9DE, UK .}

\author{Subhadip Biswas}
\affiliation{Department of Physics and Astronomy, University of Sheffield, Sheffield S3 7RH, UK.}
\affiliation{Université Paris-Saclay, CNRS, Laboratoire de Physique des Solides, 91405 Orsay, France.}

\author{Rachel Meeker}
\affiliation{Department of Physics, Kansas State University, 1228 M. L. K. Jr. Drive, Manhattan, KS, 66506, USA.}

\author{Buddhapriya Chakrabarti}
\email{buddho@ksu.edu}
\affiliation{Department of Physics, Kansas State University, 1228 M. L. K. Jr. Drive, Manhattan, KS, 66506, USA.}
\affiliation{Department of Physics and Astronomy, University of Sheffield, Sheffield S3 7RH, UK.}

\date{\today}

\begin{abstract}
We develop a self-consistent free-energy framework in which membrane shape and osmotic pressure are determined simultaneously in a finite reservoir by minimizing bending elasticity and solute entropy. Solute conservation makes osmotic pressure a thermodynamic variable rather than an externally prescribed parameter, producing a nonlinear coupling between membrane mechanics and solvent entropy. This coupling modifies the classical stability condition for spherical vesicles: instability emerges from global free-energy competition rather than the linear Helfrich stability criterion. The resulting critical pressures differ by orders of magnitude from Helfrich predictions and agree with simulations for small and large unilamellar vesicles. The framework is
relevant to cellular environments involving biomolecular condensate confinement as well as synthetic vesicles and the development of osmotic-pressure-driven encapsulation platforms.
\end{abstract}

\maketitle 

Biological membranes are fundamental structural and functional elements of living systems\cite{b:Alberts2015,b:Lipowsky1995A,p:Singer1972,p:Phillips2009}, providing compartmentalization\cite{p:Rothman1994,p:Misteli2001,p:Lingwood2010}, mechanical integrity\cite{p:Evans1990,p:Sheetz1996,p:Parsegian1981} and regulated exchange with the environment\cite{p:Hille2001,p:White1999,p:Roux2005}. From simple lipid bilayers to complex protein-rich cellular envelopes\cite{p:Helfrich1973,p:Engelman2005,p:Casares2019}, membranes must maintain stability while remaining sufficiently flexible to accommodate growth\cite{p:Gov2004,p:Sens2006,p:Yan2024}, division\cite{p:Osawa2008,p:Zeks2002,b:Alberts2015}, and environmental perturbations\cite{p:Staykova2013,p:Zilman2002,p:Lamparter2020}. 

In lower life forms such as bacteria, archaea, and simple eukaryotes, membranes operate under large and fluctuating osmotic gradients that strongly constrain mechanical stability~\cite{p:Knop2023,p:Wood2011}. Osmotic pressure differences across the membrane generate stresses that can dominate shape, tension, and integrity, driving swelling, lysis, and morphological transitions~\cite{p:Hill2002,p:Sukharev2010,p:Deng2011,p:Staykova2013}. In bacteria, survival under hypo-osmotic shock relies on tension-activated mechanosensitive channels that release osmolytes to prevent rupture~\cite{p:Sukharev2010,p:Booth2007}, while broader osmoregulatory strategies tune turgor through solute accumulation and release~\cite{p:Wood2011,p:Welsh2000}. Similar coupling between water activity, membrane mechanics, and volume regulation governs osmoadaptation and morphogenesis in yeast and other eukaryotes, underscoring the need for a quantitative framework linking osmotic forces and membrane elasticity~\cite{p:Hohmann2002,p:Knop2023,p:Evans1990}. 
\begin{figure}[b]
\includegraphics[width=\linewidth]{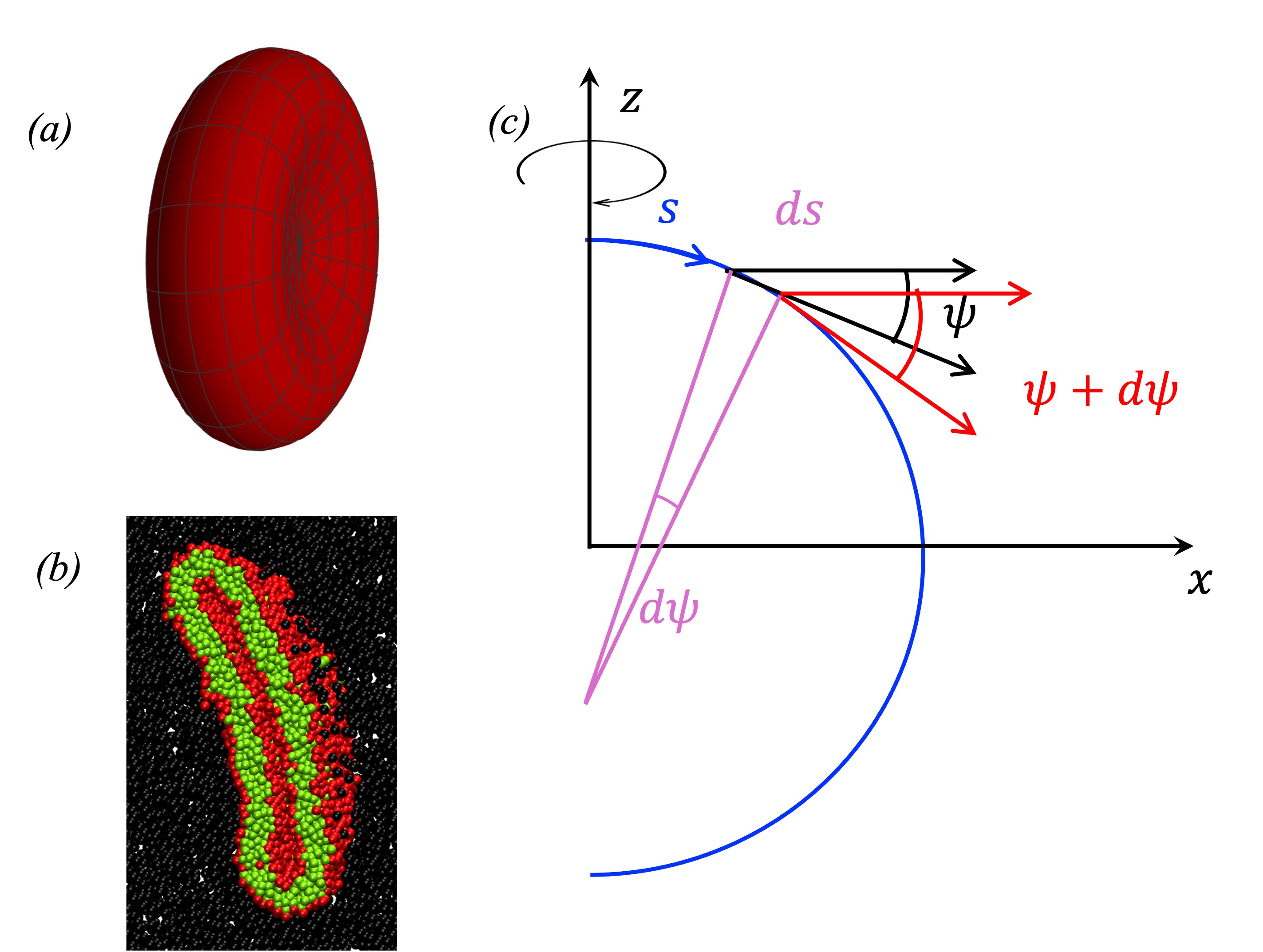}
\caption{A discotic membrane shape obtained from (a) variational shape calculation, and (b) CGMD simulations of lipid vesicles with hydrophilic head (red) and hydrophobic tail (green) groups in the presence of osmolytes (black). Panel (c) shows an axisymmetric shape in the $z(s), x(s)$ plane and the geometric parametrisation used to compute equilibrium vesicle configurations.}
\label{fig:SchematicAxisymmetric} 
\end{figure}
The Helfrich-Canham free energy~\cite{p:Canham1970,p:Helfrich1973,b:Lipowsky1995A,b:Lipowsky1995B,b:Safran1994,b:Nelson2004} has been extensively used to compute the vesicle shapes and shape transformations~\cite{p:Deuling1976,p:Lipowsky1991b,p:Zeks1997,p:Seifert1997}. These models use~\cite{p:Deuling1976,p:Helfrich1975,p:CanZhang1987,p:Julicher1994} a curvature free energy of the form
\begin{eqnarray}
 F_{H} &=& \int \left[ \frac{1}{2} k_c \left( C_{1} + C_{2} - C_0 \right)^{2} + \frac{1}{2} \bar{k_{c}} C_{1} C_{2} \right] dA \nonumber \\ &+& \int \Sigma dA + \int P dV,\label{eq:Helfrich}
\end{eqnarray}
where $C_1$ and $C_2$ are the principal curvatures, $C_0$ is the spontaneous curvature, and $k_c$ and $\bar{k}_c$ are the bending rigidities associated with mean and Gaussian curvature, respectively. The last two terms impose fixed area and volume constraints via Lagrange multipliers $\Sigma$ (surface tension) and $P$ (pressure difference)~\cite{p:Deuling1976,p:Helfrich1975,p:Helfrich1976,p:Zeks1997,p:Seifert1997,p:Julicher1994}. For closed vesicles the Gaussian curvature energy integrates to a constant via the Gauss-Bonnet theorem~\cite{b:Nelson2004,r:Deserno2015} and can be ignored when obtaining the equilibrium shapes via a variational calculation. The problem of determining equilibrium configurations of vesicles has a long history. Deuling et al.~\cite{p:Deuling1976,p:Helfrich1975} categorized axisymmetric vesicle shapes based on the first variation of the Helfrich free energy in Eq.~\ref{eq:Helfrich}. However, the parameterization used, in which the arc length is measured from the north pole $s=0$ and non-dimensinalized via a scale factor derived from the surface area of a sphere $\ell = \sqrt{4\pi R^2_0}$ has severe limitations including singularities at the equator $s=0.5$ and the south pole $s=1$, rendering integrating the non-linear shape solutions difficult~\cite{p:Zia1991,t:Miao1992, p:Gekle2017}. Instead, a different arclength parameterisation~\cite{p:Peterson1985,p:Julicher1994,p:Bunuelos2021,p:Vlahovska2024} removes the singularity at the equator, enabling an accurate determination of shapes. For multivalued membrane configurations other parametrisations~\cite{p:Rangamani2020} have been successfully employed. Irrespective of the model and parametrisation used, the $\int P dV$ term is phenomenologically ascribed to the osmotic pressure acting across the leaflets. 

Osmotic pressure has long been recognized as a key control parameter for vesicle morphology within the Helfrich curvature framework, where pressure enters as a Lagrange multiplier enforcing volume constraints~\cite{p:Helfrich1973,p:Deuling1976}. Early theoretical studies showed that increasing pressure destabilizes spherical vesicles and drives shape transitions, including budding and nonaxisymmetric deformations~\cite{p:CanZhang1987,p:Seifert1997}. Subsequent work systematically mapped pressure–curvature phase diagrams for axisymmetric and nonaxisymmetric shapes in spontaneous-curvature and bilayer-couple models~\cite{p:Julicher1994,p:Seifert1991}. While these approaches successfully classify equilibrium morphologies, osmotic pressure is typically treated phenomenologically rather than derived from solute thermodynamics, motivating recent efforts to couple membrane elasticity directly to osmolyte exchange and osmotic stresses~\cite{p:Staykova2013,p:Stenhammar2022,p:Sparr2022,p:Podgornik2023,p:Imai2024}.

A spherical vesicle minimizes Eq.~\ref{eq:Helfrich} but becomes unstable beyond a critical pressure $\Delta p_c = \frac{2 k_c}{R_0^{3}} \left(6 - C_0 R_0\right)$, where $R_0$ is the vesicle radius~\cite{p:Deuling1976}. Recent experiments on GUVs in hypotonic solutions, however, report critical pressures that exceed this prediction by up to six orders of magnitude~\cite{p:Stenhammar2022, p:Sparr2022}. Motivated by this discrepancy, we develop a parameter-free model that couples membrane mechanics to the thermodynamics of solvent exchange. In this framework, osmotic pressure emerges self-consistently from solute concentration differences across the membrane rather than being imposed phenomenologically. We compute equilibrium shapes and phase diagrams, compare them with predictions of purely mechanical bilayer models, and validate our analytical results using coarse-grained molecular dynamics (CGMD) simulations~\cite{p:Cooke2005a,p:Cooke2005b} performed via the LAMMPS package~\cite{p:Plimpton1995}.

Our starting point is the free energy
\begin{align}\label{eq:FreeEnergy}
F_{T} = F_B + \tilde{\lambda} \left( \int dA - A_0 \right) + \frac{k_B T}{v_p} \widetilde{f}(\phi)(V - V_I),
\end{align}
where $F_B$ corresponds to the bending energy term in Eq.~\ref{eq:Helfrich} (first term), and the second term corresponds to the constant area constraint imposed by the Lagrange multiplier $\tilde{\lambda}$ representing the membrane tension. The last term in Eq.~\ref{eq:FreeEnergy} describes the solution thermodynamics using a Flory--Huggins free-energy density~\cite{b:Colby2003}
\begin{align}\label{eq:FloryHuggins}
\tilde{f}(\phi) = \phi \ln(\phi) + (1- \phi) \ln (1-\phi) + \chi \phi (1-\phi),
\end{align}
where $v_p$ is the specific volume of the solute and solute volume fraction is
\begin{align}
\phi = \frac{N v_p}{V-V_I}.
\end{align}

In this expression $V_I$ is the volume of the enclosed vesicle, $N$ the number of solute particles, and $V$ is the total volume of the reservoir. Since solute particles do not permeate the semipermeable bilayer, we consider an ensemble in which all solute particles remain outside the vesicle. Consequently, Eq.~(\ref{eq:FreeEnergy}) contains no mixing entropy contribution proportional to the vesicle volume. Using a variational formulation, we determine closed axisymmetric vesicle shapes of fixed surface area $A_0$ that minimize the free energy~(\ref{eq:FreeEnergy}). 

Figure~\ref{fig:SchematicAxisymmetric}(c) illustrates the axisymmetric vesicle geometry with symmetry axis $z$. We parameterize the meridional contour by its arc length $s$, measured from the pole $x=0$~\cite{p:Julicher1994,p:Lipowsky1991b}, and describe the shape using the radial coordinate $x(s)$ and the tangent angle $\psi(s)$ measured relative to the $x$ axis. Using arc length parameterization $s$, the relations $\dot{x}=\cos\psi$, $\dot{z}=-\sin\psi$, $C_1=\dot{\psi}$, and $C_2=\sin\psi/x$ follow, where the derivative is w.r.t. the variable $s$. The enclosed volume of the vesicle is therefore given by $V_I = \int \pi x^2 \sin{\psi} ds$, while $A = \int 2 \pi x ds$ is the surface area. Note that the variables $x$, $\psi$, and $z$ are not all independent and are instead connected via the geometric constraints denoted above. The free energy expression in Eq.~\ref{eq:FreeEnergy} thus has an additional term $\int \tilde{\gamma}(s) (\dot{x}-\cos\psi) ds$ enforcing the geometrical constraint with $\tilde{\gamma}$ being the Lagrange multiplier. The free energy is thus extremized with respect to $x(s)$ and $\psi(s)$ treating them as independent variables. The total free energy is given by $F_T = S_0 + \frac{k_BT}{v_p} \tilde{f}(\phi)\left( V - V_I \right)$, where $S_0$ is the sum of bending and surface energies expressed in this coordinate system and denoted by,
\begin{equation}\label{eq:FreeEnergy2}
S_0=\int_{0}^{s_1}\!ds\Big[
\pi k_c x\Big(\dot{\psi}+\frac{\sin\psi}{x}-C_0\Big)^2
+2\pi\tilde{\lambda} x+\tilde{\gamma}(s)(\dot{x}-\cos\psi)
\Big].
\end{equation}
 Using the expressions for the volume fraction $\phi$ and $\tilde{f}(\phi)$ the first variation of the free energy in Eq.(\ref{eq:FreeEnergy2}) is therefore given by $\delta F_T = \delta S_0 + \Pi(\phi) \delta V_I$, with 
\begin{equation}
\Pi(\phi) =  - \frac{k_B T}{v_p} \left[  \ln\left( 1- \phi\right) + \chi \phi^2  \right].
\label{eq:OsmoticP}
\end{equation}
The equations for the contour obtained from the extremization of $F_T$ are 
\begin{align}
\dot{\psi}=&u, \label{eq:SE1} \\
\dot{u} =&   - \frac{u}{x} \cos{\psi}+\frac{\cos{\psi} \sin{\psi}}{x^2} + \frac{ \gamma}{x}\sin{\psi}  + \frac{\Pi(\phi)}{2k_c}x
\cos{\psi}, \label{eq:SE2}\\
\dot{\gamma} =&  \frac{(u-C_0 )^2 }{2} -\frac{\sin^2\psi}{2 x^2}   + \frac{\Pi(\phi)}{k_c}  x \sin{\psi} +\lambda,\label{eq:SE3}
\end{align}
and
\begin{align}
\gamma(s_1) =& 0, \label{eq:SE4}
\end{align}
where $\lambda = \tilde{\lambda}/k_c$, $\gamma = \tilde{\gamma}/(2 \pi k_c)$, and $s_1$ is the total arclength. These equations are solved subject to the boundary conditions $x(0) = 0$, $\psi(0) =0$, $x(s_1) = 0$, and $\psi(s_1) = \pi$, together with the constant area and geometry constraints (see SI). Further, we have $\gamma(0)=0$. This is not an independent constraint, and arises from the fact that the variational free energy in Eq.(~\ref{eq:FreeEnergy}) is not an explicit function of the arclength variable $s$, which ensures that the quantity analogous to the Hamiltonian $H$ is conserved~\cite{p:Lipowsky1991b,p:Vlahovska2024}. 
\begin{figure}[b]
\centering
\includegraphics[page=1,width=\linewidth]{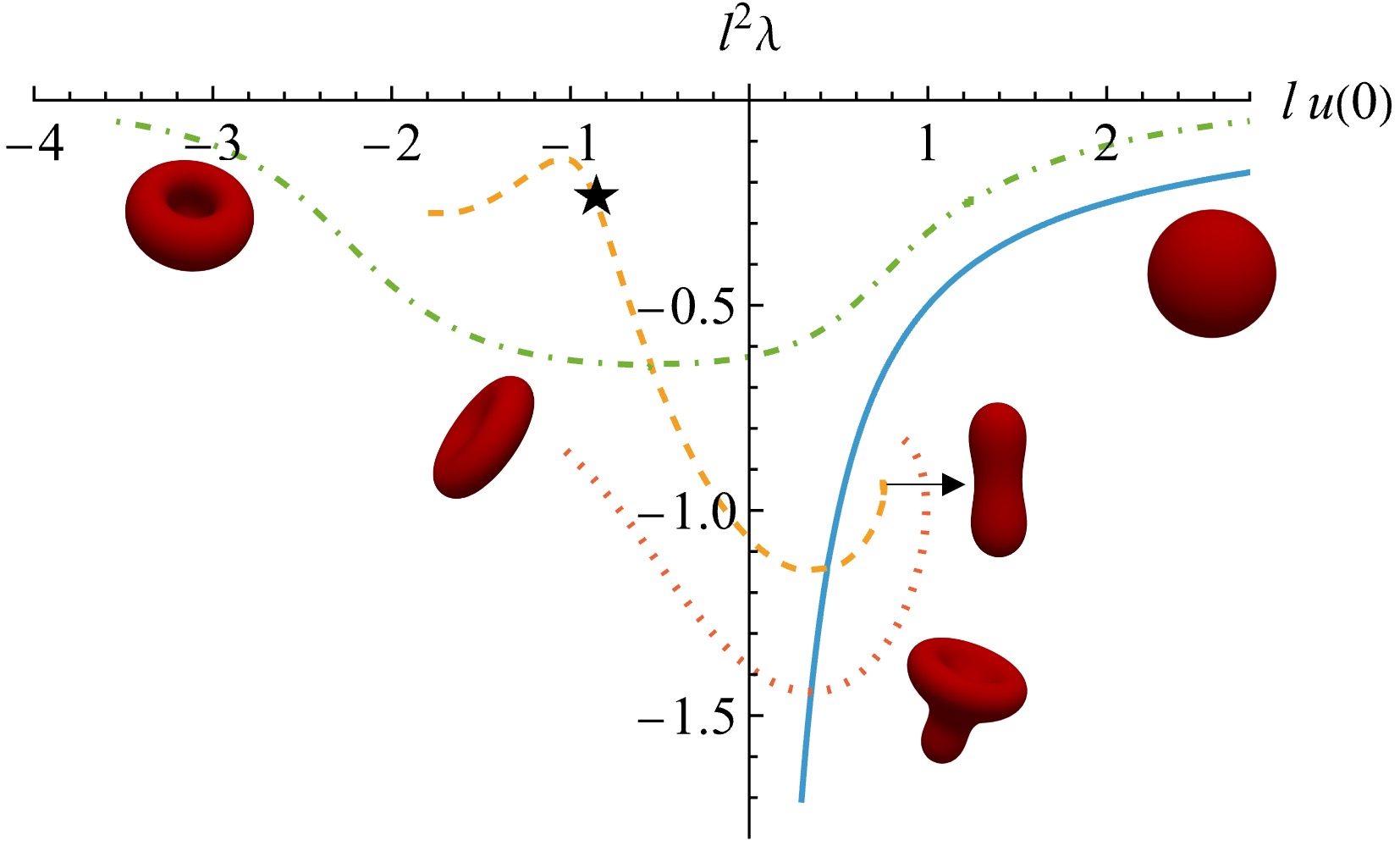}
\caption{Shape diagram in the $l u(0)-l^2 \lambda$ plane showing solutions of the scaled shape equations in the absence of osmolytes including (a) spheres ({\textcolor{teal}{blue solid}}), (b) stomatocytes ({\textcolor{ForestGreen}{green dash-dotted}}), (c) prolate and oblate shapes ({\textcolor{orange}{orange dashed}}), and (d) $L3$ shapes \textcolor{red}{red dotted}. The $\bigstar$ indicates the point on the oblate branch, such that solutions of the shape equations to its left become unphysical, self-intersecting shapes.}\label{fig:ShapeDiagramAnalytical}
\end{figure}
Eqs.(\ref{eq:SE1}-\ref{eq:SE3}) are formally analogous to Seifert's model~\cite{p:Julicher1994,p:Lipowsky1991b} with the rescaled pressure term $\bar{P}$ replaced by $\Pi(\phi)/k_c$ (see Eq.~\ref{eq:OsmoticP}). However, this difference is non-trivial and introduces an additional self-consistency condition as $\Pi(\phi)$, through $\phi$, implicitly depends on the volume enclosed by the vesicle $V_I$, which is in turn a functional of the shape $(x(s), \psi(s))$. 

\begin{figure}[b]
\centering
\includegraphics[page=1,width=\linewidth]{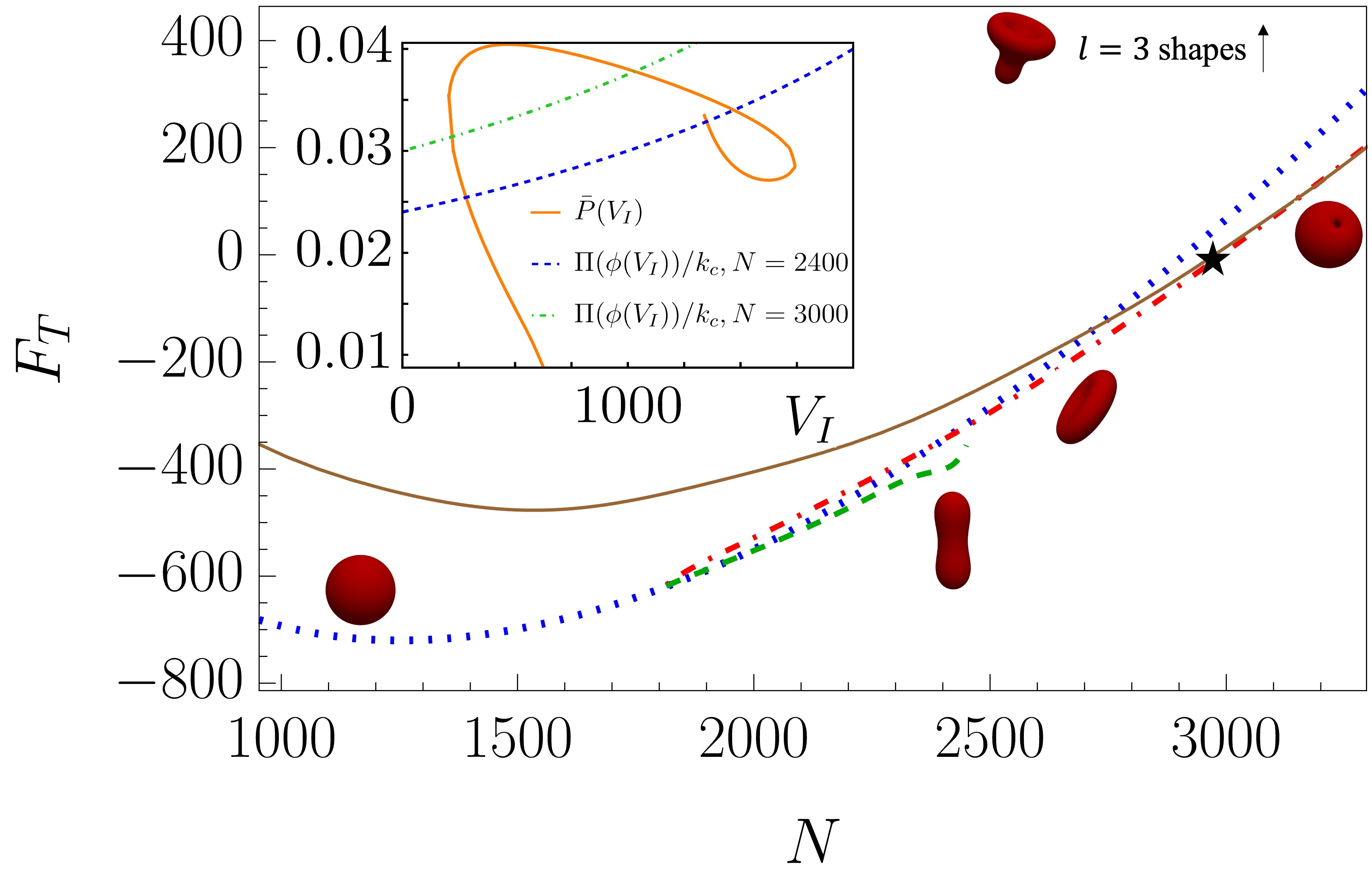}
\caption{Free energy $F_T$ (in units of $k_B T$) as a function of the osmolyte number $N$ for different shapes (a) sphere ({\textcolor{blue}{blue dots}}), (b) prolate ({\textcolor{ForestGreen}{green dashed}}), (c) oblate ({\textcolor{red}{red dash dotted}}), and (d) stomatocyte ({\textcolor{brown}{brown solid}}) showing a sequence of shapes that minimize the free energy. The free energy $F_T$ for $L3$ shapes (not shown) are much higher ($>1000 k_B T$). The $\bigstar$ indicates values of $N$ beyond which self-intersecting shapes appear in the oblate branch. Other self-intersecting solutions of higher energies are omitted. Inset: Shows $\bar{P}(V_I)$ for solutions in the oblate branch ({\textcolor{orange}{orange solid}}) and the osmotic pressure $\Pi(\phi(V_I))/k_c$ for $N=2400$ ({\textcolor{blue}{blue dashed}}) and $N=3000$ ({\g{green dash dotted}}). The intersections correspond to self-consistent solutions of the shape equations.}\label{fig:SelfConsistent}
\end{figure}

To obtain self-consistent solutions, we proceed as follows: first, we replace $\Pi(\phi)/k_c$ with $\bar{P}$ in the shape equations and scale the ensuing equations by a length scale $l =\bar{P}^{-1/3}$ to get the corresponding non-dimensionalised equations. These equations are integrated numerically to obtain the shapes for a given value of $l^2 \lambda$ as shown in Fig.~\ref{fig:ShapeDiagramAnalytical} that shows five solution branches, namely, sphere, oblate, stomatocyte, and $L3$. Next, we compute the area enclosed by the vesicle in the scaled coordinates ($A^\prime$), and use the constant area constraint, \textit{i.e.} $A= l^2 A^\prime = A_0$ to fix the pressure $\bar{P}$, which in turn uniquely fixes the volume of the vesicle $V_I = l^3 V^{\prime}_{I}$, where $V_I^\prime$ is the volume in the scaled coordinates. We thus have a set $(V_I, \bar{P})$ for each solution. These $(V_I, \bar{P})$ values are plotted together with the function $\Pi(\phi(V_I))$ for a given value of $N$, with all the other parameters fixed. See inset in Fig.~\ref{fig:SelfConsistent} for a plot of  $\Pi(\phi)$ with $(V_I, \bar{P})$ for solutions in the oblate branch. The intersections in this plot represent the solutions that respect the self-consistency condition 
\begin{equation}
\frac{\Pi(\phi(V_I))}{k_c}=\bar{P},\label{eq:self-consistency}
\end{equation}
for a given $N$.
We compare the total free energy $F_T$ of the self-consistent solutions as a function of the osmolyte number $N$. Figure~\ref{fig:SelfConsistent} shows $F_T(N)$ for distinct morphologies, revealing successive shape transitions with increasing $N$. A sphere-to-prolate transition occurs at $N \simeq 1954$, followed by a prolate-to-discocyte transition at $N \simeq 2450$. The calculations are performed in a box of volume $V \approx 5.1 \times 10^{4}\sigma^3$, with solute molecular volume $v_p \approx 0.25\sigma^3$ and miscibility parameter $\chi = 2.7$~\cite{note1}. The bending rigidity is $\kappa \approx 20k_B T$, and the vesicle area is fixed at $A_0 \approx 3300\sigma^2$. The corresponding critical osmotic pressure differences are $\Delta p_c^{SP} \approx 0.057k_B T/\sigma^{3}$ for the sphere–prolate transition and $\Delta p_c^{PO} \approx 0.066k_B T/\sigma^{3}$ for the prolate–discocyte transition. Further increase in $N$ drives discocyte flattening and yields self-intersecting configurations at $N \simeq 2990$, which remain favorable up to $N \simeq 3161$, beyond which the stomatocyte is stable~\cite{p:Munoz2014}.

The self-intersecting solutions are unphysical and reflect the absence of excluded-volume interactions in the Helfrich model; their detailed analysis is beyond the present scope. In contrast, CGMD simulations that incorporate excluded volume yield a flattened discocyte preceding the stomatocyte transition. As shown in Fig.~\ref{fig:SelfConsistent} for $N \ge 3161$, we also observe a transition from a stomatocyte to a double-membrane vesicle. However, complete closure of a stomatocyte into a double-walled vesicle lies outside our framework, as it involves a change in topology that is not captured by the theory. $L3$ shape solutions are observed for $F_T \gtrsim 1000 \;k_B T$ and is not an energy minimum. This high-energy branch compresses the scale and masks the features of interest and is not shown in Fig.~\ref{fig:SelfConsistent}. 

We employ Langevin dynamics simulations to investigate the response of a coarse-grained lipid vesicle subjected to osmotic stress. Membrane vesicles are modeled using  the Cooke--Kremer--Deserno model~\cite{p:Cooke2005a, p:Cooke2005b} with a three-bead representation of a lipid molecule consisting of one hydrophilic head bead followed by two hydrophobic tail beads performed using the LAMMPS~\cite{p:Plimpton1995} package (see SI). To impose an osmotic imbalance across the membrane, we introduce a third type of coarse-grained particle exterior to the closed vesicle. These particles have diameter $\sigma = 1.0$ and interact with all membrane beads (both head and tail) exclusively through the repulsive WCA potential, thus ensuring that the osmolytes do not stick or penetrate the membrane, and the concentration differences generate an osmotic pressure acting across the bilayer. 
\begin{figure}[h]
\centering
\includegraphics[page=1,width=\linewidth]{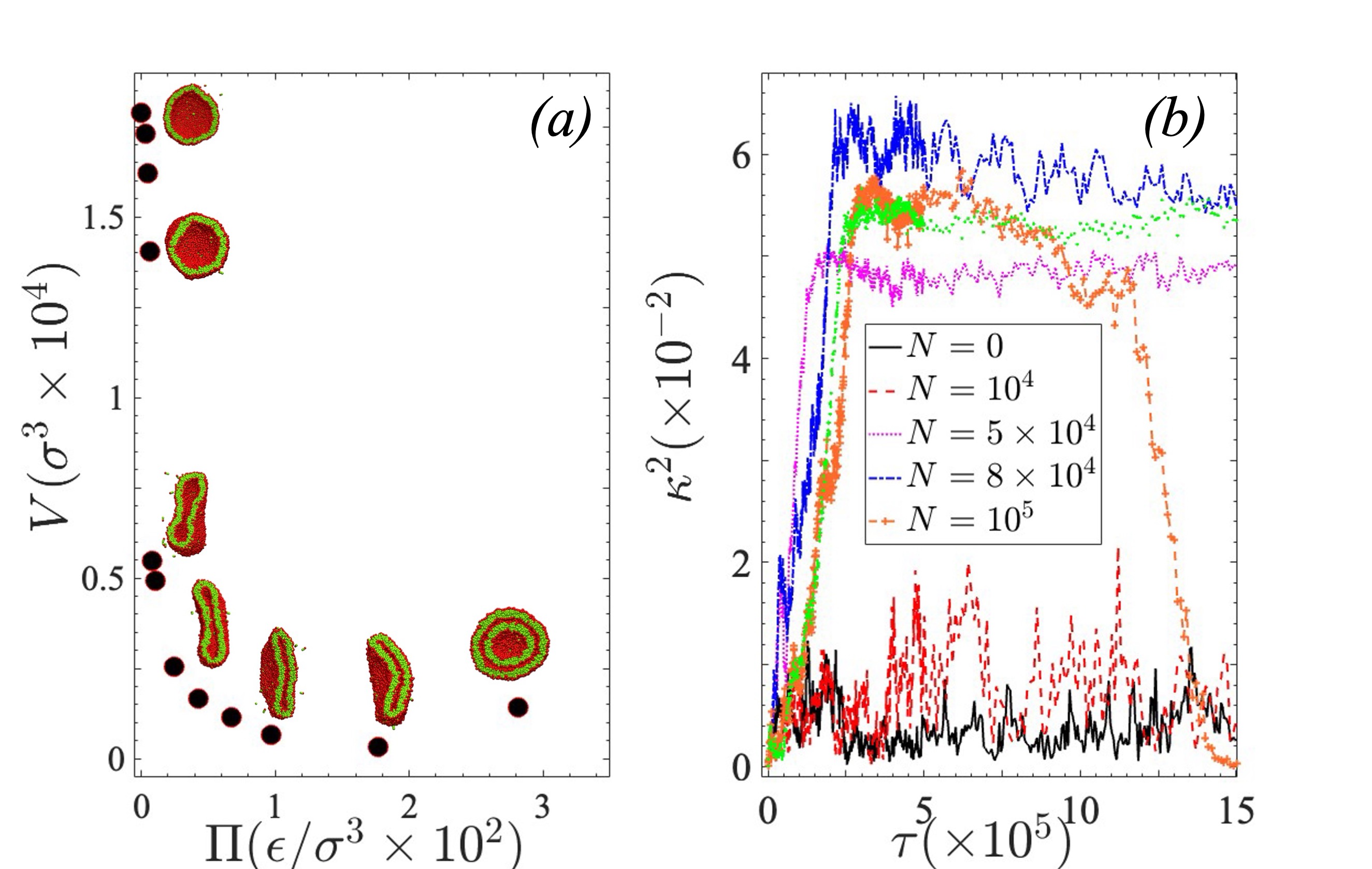}
\caption{A phase diagram obtained from CGMD simulations in the $\Pi-V$ plane, and the associated equilibrium vesicle shapes (panel (a)). The critical concentration at which a spherical vesicle becomes unstable is identified by calculating the asphericity parameter $\kappa^2$ is shown in panel (b), as a function of time $\tau$ for different osmolyte number $N$. For $N \lesssim 10^4$ $\kappa^2 \approxeq 0$, and jumps to a finite value once for $N > N_c \approx 1.75\times10^4$ the mean shape is non-spherical.} \label{fig:ShapeDiagramCGMD}
\end{figure}

Figure~\ref{fig:ShapeDiagramCGMD}(a) shows the phase diagram in the $P$--$V$ plane. The pressure is obtained from the virial expression $P=(Nk_BT+\frac{1}{3}\sum_{i<j}\mathbf{r}_{ij}\!\cdot\!\mathbf{f}_{ij})/(V-V_I)$, while the volume is computed using a voxel reconstruction approach (see SI). The vesicle volume decreases sharply with increasing osmotic pressure, and representative vesicle shapes are shown in Fig.~\ref{fig:ShapeDiagramCGMD}(a). To determine the critical pressure $\Delta p_c$ at which the spherical vesicle becomes unstable, we systematically increase the number of osmolytes and compute the relative shape anisotropy (asphericity) parameter (Fig.~\ref{fig:ShapeDiagramCGMD}(b)),
\begin{equation}\label{eq:kappasq}
\kappa^2 = \frac{3(\lambda_x^2 + \lambda_y^2 + \lambda_z^2)}{2(\lambda_x + \lambda_y + \lambda_z)^2} - \frac{1}{2},
\end{equation}
where $\lambda_i$ are the eigenvalues of the gyration tensor. For the Cooke--Deserno model, we use the parameters $\epsilon_{tt} = 1.0$ and $\epsilon_{hh} = \epsilon_{ht} = 0$, with $n_{\ell} = 8000$ lipids forming a vesicle with undeformed outer radius $R_0 \approx 25\sigma$. This corresponds to a bending modulus $\kappa = 11.97 \pm 3.4\,k_B T$ and a membrane tension $\Sigma = 0.016 \pm 0.07\,k_B T$ calculated from the fluctuation spectrum. The vesicle remains spherical for $N \lesssim 1.5\times10^4$. As the osmolyte count increases, the vesicle undergoes a sequence of shape transitions: first to a prolate shape for $1.5\times10^4 \lesssim N \lesssim 2\times10^4$, then to a discocyte for $2\times10^4 \lesssim N \lesssim 6\times10^4$, and finally to a stomatocyte for $6\times10^4 \lesssim N \lesssim 1\times10^5$. As the vesicle volume decreases with increasing osmolyte number $N$ and osmotic pressure $\Pi(c)$, the asphericity parameter $\kappa^2$ increases sharply between $N=10^4$ and $N=5\times10^4$. From this transition we estimate a critical osmolyte number $N_c \approx 1.5$--$1.75 \times 10^4$, beyond which the spherical vesicle becomes unstable (see SI). Given a simulation box of size $L=100\sigma$, the osmotic pressure for $N \approx 1.5 \times 10^{4}$ solutes in a volume $V=10^{6}\sigma^{3}$ is estimated using van’t Hoff’s law as $\Pi = k_BT (N/V) = 0.015\,k_BT/\sigma^{3}$ which shows good agreement with those obtained from our analytical calculations for similar values of the bending modulus. Taking $\sigma \simeq 0.8\,\mathrm{nm}$ as the lipid radius gives $\Pi \simeq 1.2\times10^{5}\,\mathrm{Pa} \approx 1.2\,\mathrm{bar}$. The asphericity parameter shows an interesting feature for higher osmolyte concentrations \textit{i.e.} $N\gtrsim10^5$. Here, the asphericity parameter drops to zero at sufficiently long times. This is due to an instability of the discocyte shape that leads to the formation of double-walled spherical vesicles~\cite{p:Saric2021}.

Recent experiments on osmotically stressed vesicles~\cite{p:Hallet2001,p:Zong2018,p:Dimova2020,p:Sparr2022,p:Stenhammar2022} suggests that the Helfrich--Canham description of membrane elasticity underestimates the critical pressure for vesicle instability by several orders of magnitude, even for large bending rigidities $k_c \sim 100 k_B T$. There is a fundamental disconnect between the physics operational at different scales, while the elastic theory predicts $\Delta p_c \sim k_c/R^3$, for a spherical vesicle of radius $R$, van’t Hoff’s relation $\Delta p_{\mathrm{osm}} \sim k_B T \Delta c$ operational at macroscopic scales is independent of the details of the system. It is therefore tempting to compare our results against these experiments. We note that the critical pressure obtained from the Helfrich theory~\cite{p:Deuling1976,p:Helfrich1976} is $\Delta p_c =12 k_c/R^{3}_{0} \approx 4 \times 10^{-2} \text{Pa}$ for a GUV of radius $R_0=5\;\mu m$ assuming $C_0=0$ (\textit{i.e.} bilayers), and a bending modulus of $\kappa \approx 97 k_B T$~\cite{p:Stenhammar2022, p:Sparr2022}. Using van’t Hoff’s relation this yields an equivalent solute number density $N_c=\Delta p_c/(k_BT)\simeq 10^{19}\,\mathrm{m^{-3}}$ at $T=300\,\mathrm{K}$. For CGMD simulations with an osmolyte density $\rho\simeq 0.015\,\sigma^{-3}$ with $\sigma=0.8\,\mathrm{nm}$, the corresponding number density is $n_{\mathrm{CGMD}}\simeq 3\times10^{25}\,\mathrm{m^{-3}}$. We note that extrapolating our analytical and simulation results from SUV/LUV systems to GUVs\cite{p:Sparr2022,p:Stenhammar2022} should be done with caution. The theory presented here connects the shape mechanics with solvent thermodynamics through a nonlinear mechanism. Thus, it is entirely feasible within this theory to have a non-monotonic dependence of the shape transition on the osmolyte concentration, since shape changes are accompanied by volume changes that can be significant in a finite box. This feature is qualitatively different from the classical Helfrich theory where the pressure $P$ is an external parameter that increases monotonically, rendering the spherical vesicle unstable beyond a critical threshold.

We present a framework that couples solvent thermodynamics and membrane mechanics to derive a nonlinear instability criterion for spherical vesicles, resolving this discrepancy. Previous work has examined critical osmotic swelling~\cite{p:Diamant2008,p:Diamant2012}, thermodynamic formulations of fluid vesicles~\cite{p:Diamant2011}, and elastic shape phase diagrams in the area--volume ensemble~\cite{p:Wortis1997}. However, these approaches do not determine osmotic
pressure self-consistently from solvent thermodynamics coupled to membrane mechanics, which is the focus of the present work.

Although experiments on osmotically shocked SUV/LUV systems are limited, the magnitude of the effect, governed by the balance between solvent thermodynamics and membrane bending energy, may remain significant in smaller systems near imaging resolution. Such finite-size effects may be particularly relevant when the size of a membrane compartment is only a few times larger than the enclosed object, a situation common in intracellular organelles and condensate--membrane systems. The framework developed here is therefore expected to be relevant to cellular environments, where biomolecular condensates such as the nucleolus or cytoplasmic RNA--protein droplets can mechanically interact with and reshape membrane-bound compartments under confinement and crowding conditions~\cite{p:Baumgart2023,p:Mangiarotti2023,p:Lipowsky2023}. The thermodynamic framework presented here provides a natural starting point for quantitative investigations of such condensate--membrane interactions.


{\textit{Author Contributions:}}Ideation: BC; Conceptualization: BM, RGP, and BC; Analytical Calculations: RGP, BM, MD, RM, and BC; Numerical Simulations: SG, SB, and BC; Drafting: BM, RGP, SG, RM, and BC; Editing: ALL; Funding: BC.

{\textit{Acknowledgements:}} BC thanks the Isaac Newton Institute for Mathematical Sciences, Cambridge, for support and hospitality during the SPL program, where a part of the work on this paper was carried out. This work was supported by EPSRC grant no EP/Z000580/1.

\section{Appendix}

\section{Model}
The starting point of the outlined method is the Helfrich Hamiltonian\cite{p:Helfrich1973} that models the phospholipid bilayer bounding the vesicle as a fluid membrane incorporating elastic deformations due to bending. 
\begin{eqnarray}
 F_H &=& \int \left[ \frac{1}{2} k_c \left(C_{1} + C_{2} - C_0 \right)^{2} + \frac{1}{2} \bar{k}_{c} C_1 C_2 \right] dS \nonumber \\ &+& \int \Sigma dS + \int P dV.\label{eq:Helfrich1}  
\end{eqnarray}
In Eq.~\ref{eq:Helfrich1} $C_1$, and $C_2$ are the two principal curvatures, $k_c$, and $\bar{k}_{c}$ are the bending moduli, while $\gamma$ is the interfacial tension, and $P = P_{ext} - P_{int}$ is the pressure difference acting across the membrane. The elastic deformation energy contains two terms, one proportional to the mean curvature and the other corresponding to the Gaussian curvature. For closed surfaces, the elastic deformation energy due to the Gaussian curvature integrates to a constant term on account of the Gauss-Bonnet theorem. It is henceforth ignored in the variational calculation of vesicle shapes that are minimizers of the free energy shown in Eq.~\ref{eq:Helfrich1}. 

A limitation in this framework is that the pressure difference $\Delta p$ is merely ascribed as the osmotic pressure\cite{p:Helfrich1976} instead of being calculated within the model. This limits the possibility to relate the imposed osmotic pressure to concentration differences acting across the leaflets. To remove this limitation we minimise the total energy of the system composed of the Helfrich free energy in Eq.~\ref{eq:Helfrich1} and a mixing free energy that dictates the partitioning of the solute across the semi-permeable membrane. The total free energy of the system is given by
\begin{align}\label{eq:FreeEnergy}
F_{T} = F_{B} + F_{mix},
\end{align}
where $F_{mix}$ is the mixing free energy. The mixing free energy is expressed as $F_{mix} = \int \tilde{f}[\phi] dV$, with the free energy density $\tilde{f}[\phi]$ assumed to have the Flory-Huggins form\cite{b:Colby2003}
\begin{align}\label{eq:Flory-Huggins}
\tilde{f}[\phi] = \frac{F[\phi]}{k_B T} = \phi \ln \phi + (1 - \phi) \ln (1-\phi) + \chi \phi (1-\phi).
\end{align}
Here, $\phi$ refers to the volume fraction of the solute and $\chi$ the miscibility parameter that dictates the theremodynamic equilibrium between the solute and solvent. The volume fraction is given by 
\begin{align}\label{eq:PhiDefinition}
\phi = \frac{N v_p}{V-V_I}.
\end{align}
where $N$ is the number of solute particles outside the vesicle, $v_p$ the volume occupied by a solute molecule, $V$ the total volume of the system, and $V_I$ the volume enclosed by the vesicle. We assume an initial conformation where the vesicle is impermeable to the solute. The total energy of the vesicle to be minimized is therefore given by 
\begin{align}\label{eq:OsmoticVesicleFreeEnergy}
F_T = F_B+ \frac{k_BT}{v_p} \widetilde{f}(\phi)(V - V_I). 
\end{align}
Our objective is thus to find the closed shape(s) that minimize the free energy~(\ref{eq:OsmoticVesicleFreeEnergy}) for a fixed area $A_0$, restricting ourselves to axisymmetric shapes. 

\section{Geometric parametrization}
\begin{figure}
\includegraphics[width=0.5\linewidth]{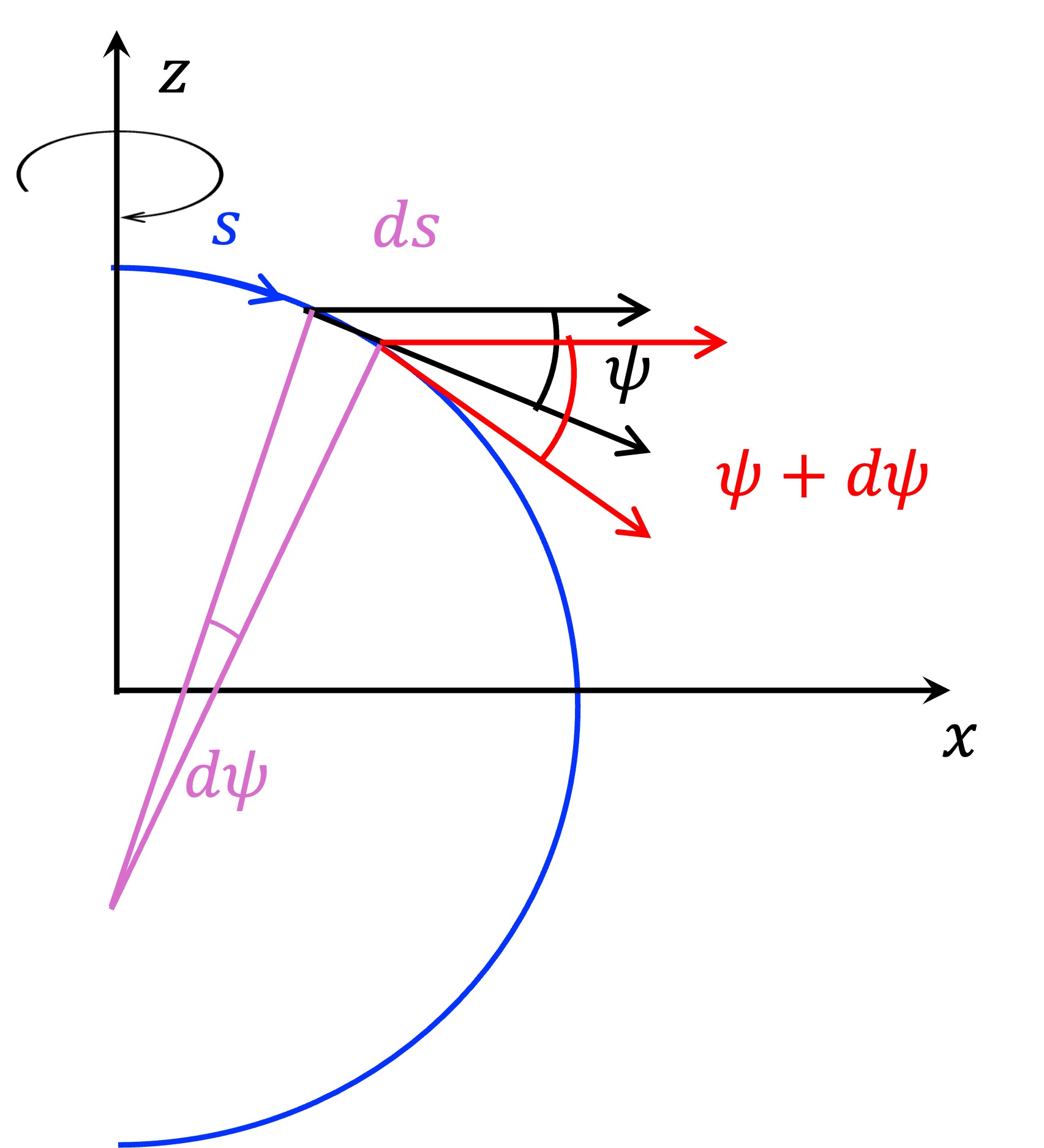}
\caption{
Schematic of an axisymmetric vesicle shape. 
The $z$-axis denotes the axis of symmetry, while $s$ is the arclength along the meridional curve $(x(s),z(s))$. The angle $\psi(s)$ between the radial direction and the tangent to the curve varies from $0$ to $\pi$ between the two poles.
}
\label{fig:AxiymmetricShapeSI}
\end{figure}

We consider axisymmetric shapes as shown in Fig.~\ref{fig:AxiymmetricShapeSI} with the $z$-axis chosen as the symmetry axis, and $x$ is an axis perpendicular to $z$. We parameterize the contour that defines the edge of the shape in the $x$-$z$ plane using  the arclength $s$ along the contour starting from $x =0$, and use $x(s)$ and the angle  $\psi(s)$ substended by the tangent to the contour on the $x$ axis as the coordinates~\cite{p:Julicher1994}. 
The following geometric relations can be easily derived.
\begin{align}
\dot{x}  &= \cos{\psi} \label{eq:xpsiconstraint} \\
\dot{z} &= - \sin{\psi} \\
C_1 &= \dot{\psi}  \\
C_2& = \sin{\psi}/x,
\end{align}
where a dot over the symbols indicate derivative with respect to $s$. The free energy Eq.~(\ref{eq:OsmoticVesicleFreeEnergy}) in these coordinates has the form
\begin{align}
\label{eq:VesicleFreeEnergy}
F_{T} = \int_0^{s_1} \! \!  ds \, \pi k_c x \left( \dot{\psi} + \frac{\sin \psi}{x} - C_0 \right)^2 + \frac{k_B T}{v_p} \tilde{f}(\phi) \left( V -V_I \right)
\end{align}
with 
\begin{align}
\label{Volume}
V_I = \int_0^{s_1} ds \, \pi x(s)^2 \sin \psi(s),
\end{align}
being the volume of the vesicle. The coordinates $x(s)$ and $\psi(s)$ are not independent but related through Eq.~(\ref{eq:xpsiconstraint}). Further, we consider equilibrium shapes having a fixed area  $A_0$. Incorporating these constraints into the free energy $F_T$ with the help of Lagrange multipliers $\tilde{\gamma}(s)$ and $\tilde{\lambda}$ yields the quantity
\begin{widetext}
\begin{align}\label{eq:SeifertParametrisation}
F_T = \int_0^{s_1} \! \!  ds \, \pi k_c x \left( \dot{\psi} + \frac{\sin \psi}{x} - C_0 \right)^2 + \frac{k_B T}{v_p} \tilde{f}(\phi) \left( V -V_I \right) +  \int_0^{s_1} \! \! ds\, \tilde{\gamma}(s) \left( \dot{x} - \cos{\psi} \right) + {\tilde{\lambda}} \left( 2 \pi \int_0^{s_1} \! \! ds \, x - A_0 \right) ,
\end{align}
which we extremize with respect to $x(s)$ and $\psi(s)$ to obtain the equations for the equilibrium shapes.  For convenience, we rewrite Eq.~(\ref{eq:SeifertParametrisation}) as
\begin{align}
F_T= S_0 + \frac{k_B T}{v_p} \widetilde{f}(\phi)(V - V_I)
\end{align} 
where 
\begin{align}\label{S0}
S_0 = \int_0^{s_1} \! \!  ds \, \left[ \pi k_c x \left( \dot{\psi} + \frac{\sin{\psi}}{x} - C_0 \right)^2 + 2 \pi \tilde{\lambda} x + \tilde{\gamma}(s) \left( \dot{x} - \cos{\psi} \right) \right].
\end{align} 
\end{widetext}
Carrying out the first variation we obtain,
\begin{align}\label{eq:FirstVariation}
\delta F_T = \delta S_0 + \frac{k_B T}{v_p} \left[ -\widetilde{f}(\phi) \delta V_I + (V - V_I) \delta \widetilde{f}(\phi) \right].
\end{align}
Using the Flory-Huggins free energy Eq.~(\ref{eq:Flory-Huggins}) and the expression for $\phi$ Eq.(\ref{eq:PhiDefinition}) we obtain,
\begin{align}\label{eq:FirstVariationFloryHuggins}
\delta \widetilde{f} = \frac{\phi}{V - V_I} \left[ \ln{\left(\frac{\phi}{1-\phi} \right)}+ \chi (1-2 \phi )  \right] \delta V_I.
\end{align}
Substituting this in Eq.~(\ref{eq:FirstVariation}) yields
\begin{align}\label{deltaS}
\delta F_T = \delta S_0 + \Pi(\phi) \delta V_I
\end{align}
where 
\begin{align}
\Pi(\phi) =  - \frac{k_B T}{v_p} \left[  \ln\left( 1- \phi\right) + \chi \phi^2  \right]
\end{align}
is the difference in the osmotic pressures across the membrane. On explicitly carrying out the variation in Eq.~(\ref{deltaS}), subject to the boundary conditions
\begin{align}
x(0) =& 0\label{bcFinal} \\
\psi(0) =&0 \\
x(s_1) = & 0  \text{ and }\\
\psi(s_1) = & \pi,
\end{align}
we obtain (see Appendix \ref{sec:variation}) the shape equations
\begin{align}
\dot{\psi}=&u \label{SE1} \\
\dot{u} =&   - \frac{u}{x} \cos{\psi}+\frac{\cos{\psi} \sin{\psi}}{x^2} + \frac{ \gamma}{x}\sin{\psi}  +  \frac{\Pi(\phi)}{2k_c}x \cos{\psi} \label{SE2}\\
 \dot{\gamma} =&  \frac{(u-C_0 )^2 }{2} -\frac{\sin^2\psi}{2 x^2}   +   \frac{\Pi(\phi)}{k_c}  x \sin{\psi} +\lambda \label{SE3} 
\end{align}
together with the boundary condition
\begin{align}\label{bcLast}
\gamma(s_1) = 0.
\end{align}
In Eq.~(\ref{SE1}) we have used $\lambda \equiv \tilde{\lambda}/k_c$ and $\gamma \equiv \tilde{\gamma}/(2 \pi k_c)$. These equations are to be solved with the constraints 
\begin{align}
\dot{x} =& \cos{\psi} \text{ and } \\
2 \pi \int_0^{s_1}\! \! ds x ds =&A_0.
\end{align}
We further have $\gamma(0) = 0$, which is not an additional constraint but automatically satisfied since $F_T$ is not a function of $s$ explicitly (see Appendix~\ref{sec:variation}). 

If we make the replacement $\Pi(\phi)/k_c \rightarrow \bar{P}$, and ignore the dependence on $\phi$, we retrieve the equations in~\cite{p:Lipowsky1991b}, which are 
\begin{align}
\dot{\psi}=&u  \label{seifert1} \\
\dot{u} =&   - \frac{u}{x} \cos{\psi}+\frac{\cos{\psi} \sin{\psi}}{x^2} + \frac{ \gamma}{x}\sin{\psi}  + \frac{\bar{P}}{2} x \cos{\psi} \\
 \dot{\gamma} =&  \frac{(u-C_0 )^2 }{2} -\frac{\sin^2\psi}{2 x^2}   + \bar{P}  x \sin{\psi} +\lambda \\
\dot{x} =& \cos{\psi}, \label{seifert2}
\end{align}
with the boundary conditions remaining the same. Note that we can non-dimensionalize these equations  by introducing a length scale 
\begin{align}\label{lp}
l= \bar{P}^{-1/3}
\end{align} 
and defining the dimensionless variables
\begin{align}
s^\prime = s/l, \; x^\prime = x/l, \; \psi^\prime = \psi, \; u^\prime = lu, \; \gamma^\prime = l \gamma, \\
C_0^\prime =l C_0, \; \lambda^\prime = l^2 \lambda. 
\end{align}
In terms of the primed variables Eqs.~(\ref{seifert1})--(\ref{seifert2}) reads as 
\begin{align}
\dot{\psi^\prime}=&u^\prime  \\
\dot{u^\prime} =&   - \frac{u^\prime}{x^\prime} \cos{\psi^\prime}+\frac{\cos{\psi^\prime} \sin{\psi^\prime}}{{x^\prime}^2} + \frac{ \gamma^\prime}{x^\prime}\sin{\psi^\prime}  + \frac{1}{2} x^\prime \cos{\psi^\prime} \\
 \dot{\gamma^\prime} =&  \frac{(u^\prime-C_0^\prime )^2 }{2} -\frac{\sin^2\psi^\prime}{2 {x^\prime}^2}   +   x^\prime \sin{\psi^\prime} +\lambda^\prime \\
\dot{x^\prime} =& \cos{\psi^\prime},
\end{align}
which is of the same form as Eq.~(\ref{seifert1})--(\ref{seifert2})  with $\bar{P} = 1$, thus showing that the shape equations~(\ref{seifert1})--(\ref{seifert2}) are scale invariant. On solving these equations for various values of $\lambda^\prime$ we obtain the shape diagram shown in Fig.~2 of main text. The scale invariance of the shape equations is exploited when we fix $l$, and therefore $\bar{P}$, by scaling the solutions meet the area constraint $A = l^2 A^\prime = A_0$.

\section{Notes on variation of $F_T$}\label{sec:variation}
Here we demonstrate how the shape equations and the boundary condtions on the geometric factor $\gamma$ are obtained from variation of the total free energy $F_T$.  For brevity we introduce the variables $y_1 \equiv x$ and $y_2 \equiv \psi$ and also adopt the convention that the index $i$ is to be summed over whenever repeated within a term. The integrand in Eq.~(3) is denoted here by 
\begin{align} \label{L}
\mathcal{L}(y_i, \dot{y}_i) = & \pi k_c y_1 \left( \dot{y}_2 + \frac{\sin y_2}{y_1} - C_0 \right)^2 + 2 \pi \tilde{\lambda} y_1 + \nonumber \\ 
&\tilde{\gamma}(s)(\dot{y_1} - \cos y_2).
\end{align}
Likewise, the integrand in the expression for volume $V_I$ is denoted by 
\begin{align}\label{LPrime}
\mathcal{L}^\prime (y_i) = \pi y_1^2 \sin y_2.
\end{align}
 A smooth axisymmetric shape demands the boundary conditions  
\begin{align}
 &y_1(0) = 0,   &y_2(0) = 0 \label{BC1} \\
 &y_1(s_1) = 0, &y_2(s_1) = \pi \label{BC2}
\end{align}  

 Suppose the countour $y_i(s)$  extending between $s=0$ and $s_1$ extremizes $F_T$ over all contours that satisfy the boundary conditions~(\ref{BC1}) and (\ref{BC2}), with the end-point $s_1$ free to vary. Then, any infinitesimal varition about this controur should render $\delta F_T =0$.   Formally, following the standard procedure in variational calculus, we introduce a new contour that is infinitismally close to $y_i(s)$:
\begin{align}\label{newPath}
\hat{y}_i(s) = y_i(s) + \epsilon \eta_i(s),
\end{align}
where $\epsilon$ is an infinitesimal parameter. The new contour is assumed to extend from $s =0$ to 
\begin{align}
\hat{s}_1 = s_1 + \epsilon s_1^\prime .
\end{align}
If $\hat{s}_1 > s_1$, then $y_i$ for $s_1<s<\hat{s}_1$ is defined by Taylor expansion of $y_i$ about $s_1$. Note that the new contour is also forced to satisfy
\begin{align}
 &\hat{y}_1(0) = 0,   &\hat{y}_2(0) = 0 \label{BBC1} \\
 &\hat{y}_1(\hat{s}_1) = 0, &\hat{y}_2(\hat{s}_1) = \pi, \label{BBC2}
\end{align}  
which ensures that the extremization is done over the subset of curves that respect these boundary conditions. As already stated in the main text, we may break up the variation of the total free energy $F_T$ into two parts as 
\begin{align}\label{DeltaFT}
\delta F_T = \delta S_0 + \Pi(\phi) \delta V_I. 
\end{align} 
We first evaluate
\begin{align}
\delta S_0 \equiv  S_0[\hat{y}_i] - S_0[y_i] = \int_0^{\hat{s}_1} \! \! ds \, \mathcal{L}(\hat{y}_i, \dot{\hat{y}}_i) - \int_0^{s_1} \! \! ds \, \mathcal{L}(y_i, \dot{y}_i).
\end{align}

 Splitting the first integral in the above immediatly yields
\begin{align}\label{variationStep}
\delta S_0 =&\int_0^{s_1} \! \! ds \, \mathcal{L}(\hat{y}_i, \dot{\hat{y}}_i) +\nonumber \\
&\int_{s_1}^{s_1+\epsilon s_1^\prime} \! \! ds \, \mathcal{L}(\hat{y}_i, \dot{\hat{y}}_i) - \int_0^{s_1} \! \! ds \, \mathcal{L}(y_i, \dot{y}_i),
\end{align}
\begin{widetext}
The first and the last terms are combined to obtain
\begin{align}
\int_0^{s_1} \! \! ds \, \mathcal{L}(\hat{y}_i, \dot{\hat{y}}_i) -  \int_0^{s_1} \! \! ds \, \mathcal{L}(y_i, \dot{y}_i) =& \int_0^{s_1} \! \! ds \, \frac{\partial \mathcal{L}}{\partial y_i} \delta y_i +  \int_0^{s_1} \! \! ds \, \frac{\partial \mathcal{L}}{\partial \dot{y}_i} \delta \dot{y}_i \\
=& \epsilon \int_0^{s_1} \! \! ds \, \frac{\partial \mathcal{L}}{\partial y_i} \eta_i + \epsilon \left. \frac{\partial \mathcal{L}}{\partial \dot{y}_i} \eta_i \right|_{s_1} -\epsilon  \int_0^{s_1} \! \! ds \, \left( \frac{d}{ds} \frac{\partial \mathcal{L}}{\partial \dot{y}_i} \right) \eta_i,
\end{align}
where, in the second line, we have integrated by parts and set $\eta_i(0) = 0$, as we are varying over contours with fixed initial point.  Using the above result in Eq.~(\ref{variationStep}) we get
\begin{align}\label{deltaS0}
 \delta S_0 =\epsilon \int_0^{s_1} \! \! ds \,  \eta_i \left[\frac{\partial \mathcal{L}}{\partial y_i}  - \left( \frac{d}{ds} \frac{\partial \mathcal{L}}{\partial \dot{y}_i} \right)\right]+ \epsilon \left. \left( \frac{\partial \mathcal{L}}{\partial \dot{y}_i} \eta_i + s_1^\prime \mathcal{L} \right) \right|_{s_1} 
\end{align}
to $\mathcal{O}(\epsilon)$. From Eq.~(\ref{newPath})  we obtain  
\begin{align}\label{dely}
\hat{y}_i(\hat{s}_1) - y_i(s_1) =  \epsilon \left( \eta_i(s_1) + s_1^\prime \dot{y}_i (s_1) \right) 
\end{align}
by expanding $y_i$ and $\eta_i$ about $s=s_1$ and retaining terms to $\mathcal{O}(\epsilon)$. 
Since the left hand side in Eq.~(\ref{dely}) evaluates to $0$ by virtue of Eqs.~(\ref{BC2}) and (\ref{BBC2}), we have $\eta_i(s_1)=-s_1^\prime \dot{y}_i(s_1)$, using which in Eq.~(\ref{deltaS0}) immediatly renders it  as 
\begin{align}\label{DeltaS0}
\delta S_0 =\epsilon \int_0^{s_1} \! \! ds \,  \eta_i \left[\frac{\partial \mathcal{L}}{\partial y_i}  - \left( \frac{d}{ds} \frac{\partial \mathcal{L}}{\partial \dot{y}_i} \right)\right] - s_1^\prime \epsilon \left. \left( \frac{\partial \mathcal{L}}{\partial \dot{y}_i} \dot{y}_i -\mathcal{L} \right) \right|_{s_1} 
\end{align} 
Note that in obtaining this expression for $\delta S_0$ we have not used the explicit form of $\mathcal{L}$ anywhere. Hence, the expression for $\delta V_I$ may be obtained simply by replacing $\mathcal{L}$ in the in Eq.~(\ref{DeltaS0}) with $\mathcal{L}^\prime$, yielding 
\begin{align}\label{DeltaVI}
\delta V_I =\epsilon \int_0^{s_1} \! \! ds \,  \eta_i \left[\frac{\partial \mathcal{L}^\prime}{\partial y_i}  - \left( \frac{d}{ds} \frac{\partial \mathcal{L}^\prime}{\partial \dot{y}_i} \right)\right] - s_1^\prime \epsilon \left. \left( \frac{\partial \mathcal{L}^\prime}{\partial \dot{y}_i} \dot{y}_i -\mathcal{L}^\prime \right) \right|_{s_1}.
\end{align} 
For generality, we have kept the terms involving the derivative with respect to $\dot{y}_i$ despite $\mathcal{L}^\prime$ not being a  function of $\dot{y}_i$. 
Substituting for $\delta S_0$ and $\delta V_I$ in Eq.~(\ref{DeltaFT}) using Eqs.~(\ref{DeltaS0}) and (\ref{DeltaVI}), we obtain

\begin{align}
\delta F_T =&\epsilon \int_0^{s_1} \! \! ds \,  \eta_i \left[\frac{\partial \mathcal{L}}{\partial y_i}  - \left( \frac{d}{ds} \frac{\partial \mathcal{L}}{\partial \dot{y}_i} \right) + \Pi(\phi)\left\{ \frac{\partial \mathcal{L^\prime}}{\partial y_i}  - \left( \frac{d}{ds} \frac{\partial \mathcal{L^\prime}}{\partial \dot{y}_i} \right)\right\} \right]
 \nonumber \\
&-s_1^\prime \epsilon \left[ \frac{\partial \mathcal{L}}{\partial \dot{y}_i} \dot{y}_i -\mathcal{L} +\Pi(\phi) \left\{  \frac{\partial \mathcal{L^\prime}}{\partial \dot{y}_i} \dot{y}_i -\mathcal{L^\prime}  \right\}\right]_{s_1}, 
\end{align}
to $\mathcal{O}(\epsilon)$.  At the extremal contour, $\delta F_T =0$ to  $\mathcal{O}(\epsilon)$ for any choice of $\eta_i(s)$ and $s_1^\prime$. This implies that the integrand in the first line and the term inside the bracket in the second line evaluated at $s_1$ should both be $0$ independently. Further, for the same reason, the integrand in the first line should be $0$ for $i =1$ and $2$ independently.  Explicitly, 
\begin{align}
\frac{\partial \mathcal{L}}{\partial y_i}  - \left( \frac{d}{ds} \frac{\partial \mathcal{L}}{\partial \dot{y}_i} \right) + \Pi(\phi)\left\{ \frac{\partial \mathcal{L^\prime}}{\partial y_i}  - \left( \frac{d}{ds} \frac{\partial \mathcal{L^\prime}}{\partial \dot{y}_i} \right)\right\} = 0, \text{ for } i =1,2 \label{SEA1} \\
 \left[ \frac{\partial \mathcal{L}}{\partial \dot{y}_i} \dot{y}_i -\mathcal{L} +\Pi(\phi) \left\{  \frac{\partial \mathcal{L^\prime}}{\partial \dot{y}_i} \dot{y}_i -\mathcal{L^\prime}  \right\}\right]_{s_1} =0 \label{SEA2}
\end{align}

\end{widetext}
Equation~(\ref{SEA1}), together with Eqs.~(\ref{L}) and (\ref{LPrime}), yields the shape equations~(9) and (8) in the main text for $i =1$ and $2$, respectively. Likewise, Eq.~(\ref{SEA2}) yields the boundary condition $\gamma(s_1) =0$. 

 To capitalize from the fact that neither $\mathcal{L}$ nor $\mathcal{L}^\prime$ is a function of $s$ explicitely, we define the quantity 
\begin{align}\label{Ham}
H(s) = \frac{\partial \mathcal{L}}{\partial \dot{y}_i} \dot{y}_i -\mathcal{L} +\Pi(\phi) \left\{  \frac{\partial \mathcal{L^\prime}}{\partial \dot{y}_i} \dot{y}_i -\mathcal{L^\prime}  \right\}.
\end{align}
which is analogous to the Hamiltonian. It is easily varified that $dH/ds =0$ along an extremal contour, and $H(s_1) = 0$ by virtue of Eq.~(\ref{SEA2}). It follows that 
\begin{align}
H(0) = \gamma(0) = 0,
\end{align}
where the first equality comes from substituting for $\mathcal{L}$ and $\mathcal{L}^\prime$ in Eq.~(\ref{Ham}) using Eqs.~(\ref{L}) and (\ref{LPrime}) and then evaluating the ensuing expression at $s=0$ with the help of the boundary condtions~(\ref{BC1}).

\section{CGMD simulations}
\subsection{CG model of spherical vesicle under osmotic stress}

We employ Langevin dynamics simulations to investigate the response of a coarse-grained lipid vesicle subjected to osmotic stress. Membrane vesicles are modeled using a coarse-grained lipid bilayer framework based on the Cooke--Kremer--Deserno model \cite{p:Cooke2005a, p:Cooke2005b}. In this three-bead representation, each lipid molecule consists of one hydrophilic head bead followed by two hydrophobic tail beads.

Non-bonded interactions between coarse-grained beads are modeled using either a purely repulsive Weeks--Chandler--Andersen (WCA) potential or a truncated Lennard--Jones (LJ) potential. The repulsive interaction is given by
\begin{equation}
V_{\mathrm{WCA}}(r) =
\begin{cases}
4\epsilon \left[ \left( \dfrac{\sigma}{r} \right)^{12}
- \left( \dfrac{\sigma}{r} \right)^6 + \dfrac{1}{4} \right], & r \le r_c, \\
0, & r > r_c,
\end{cases}
\label{eq:Vrep}
\end{equation}
where $\epsilon$ denotes the interaction strength and the cutoff distance $r_c = 2^{1/6}\sigma$ corresponds to the position of the potential minimum. 

The geometric parameters of the lipid beads are chosen as $\sigma_{\mathrm{head,head}} = \sigma_{\mathrm{head,tail}} = 0.95\sigma$ and $\sigma_{\mathrm{tail,tail}} = \sigma$, such that tail beads are slightly larger than head beads. To reproduce the fluid nature of the membrane, additional nonbonded interactions between tail beads are introduced. These effective hydrophobic interactions, which mimic the presence of an implicit solvent, are described by an attractive potential of the form
\begin{equation}
V_{\mathrm{cos}}(r) =
\begin{cases}
-\epsilon, & r < r_c, \\
-\epsilon \cos^2 \left[ \dfrac{\pi (r - r_c)}{2 w_c} \right], & r_c \le r \le r_c + w_c, \\
0, & r > r_c + w_c,
\end{cases}
\label{eq:Vcos}
\end{equation}
where $w_c$ defines the width of the attractive region and the potential smoothly decays to zero over the interval $[r_c, r_c + w_c]$.

\begin{figure}[!b]
\includegraphics[width=0.65\linewidth]{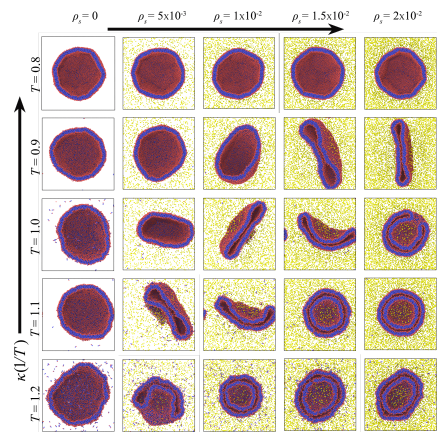}
\caption{Temperature-osmotic pressure phase diagram of vesicle deformation in equilibrium. CGMD cross-sectional snapshots of a spherical lipid bilayer vesicle ({\textcolor{red}{red}} being head and {\textcolor{blue}{blue}} tail beads) with initial radius $R_0 = 25\sigma$ subjected to changing in external solute (yellow beads) density $\rho_s$ ranging from $0$ to $2\times 10^{-2} \sigma^{-3}$ keeping $w_c=1.4$ fixed. Temperature is varied from $T = 0.8\varepsilon/k_B$ to $1.2\varepsilon/k_B$, effectively decreasing the bending rigidity. In the absence of $\rho_s$, serve as the reference state, spherical vesicles are mostly undeformed. A similar aspect also emerges at higher membrane bending rigidity or at low temperature. There are deformed and double-vesicle morphologies at higher $T$ and $\rho_s$.}
\label{fig:deformation_phase_diagram}
\end{figure}

To impose an osmotic imbalance across the membrane, additional coarse-grained particles of type~3 are introduced and act as osmolytes. These particles have diameter $\sigma = 1.0$ and interact with all membrane beads (both head and tail) exclusively through the repulsive WCA potential, ensuring excluded-volume interactions without adhesion to or penetration of the membrane. Differences in osmolyte concentration inside and outside the vesicle generate an effective osmotic pressure acting on the membrane.

Bonded interactions between adjacent beads along a lipid molecule are modeled using the finitely extensible nonlinear elastic (FENE) potential,
\begin{equation}
V_{\mathrm{FENE}}(r) = -\dfrac{k r_0^2}{2}
\ln \left[ 1 - \left( \dfrac{r}{r_0} \right)^2 \right],
\label{eq:VFENE}
\end{equation}
where $k$ is the spring constant and $r_0$ denotes the maximum bond extension.

To enforce a preferred linear conformation of lipid molecules, a bond-angle potential is applied,
\begin{equation}
V_{\mathrm{angle}} = \dfrac{k_\theta}{2} (\theta - \theta_0)^2,
\label{eq:Vangle}
\end{equation}
with equilibrium angle $\theta_0 = 180^\circ$ and bending stiffness $k_\theta = 10\epsilon/\sigma^2$.

All molecular dynamics simulations were performed using Large-scale Atomic/Molecular Massively Parallel Simulator (LAMMPS) \cite{p:Plimpton1995}

\subsection{ Osmotic stress - membrane rigidity phase diagram of vesicle deformation:}

To investigate membrane mechanics under osmotic stress, we performed coarse-grained molecular dynamics (CGMD) simulations of lipid bilayer membrane forming an initially spherical vesicle of radius $R_0 = 25\sigma$ with different solute densities $\rho_s$ in Fig.~\ref{fig:deformation_phase_diagram}. Osmotic pressure in CG simulation is imposed by varying the external solvent particle number density $\rho_s$ from $0$ to $2\times10^{-2} \sigma^{-3}$, thereby generating a pressure difference across the membrane vesicle interior and exterior. We systematically varied temperatures in the range $T = 0.8\varepsilon/k_B$ to $1.2\varepsilon/k_B$, where increasing temperature effectively reduces the membrane bending modulus $\kappa$. The resulting phase diagram in Fig.~\ref{fig:deformation_phase_diagram} reveals that vesicle morphology is governed by the competition between osmotic compression and thermally softened bending rigidity. A similar entropic pressure-driven mechanism of membrane deformation has been observed for a confined polymer under soft membrane confinement \cite{p:Biswas2025}. Here, the spherical vesicles largely retain their initial spherical geometry even at finite osmotic stress at low temperature (high bending stiffness $\kappa$). However, as the temperature increases, surface fluctuations reduce  $\kappa$ which promotes shape instabilities including prolate, oblate and budded conformations. At sufficiently high osmotic density and high temperature, we observe membrane invagination and the spontaneous formation of double-vesicle (vesicle-within-vesicle) structures. This result demonstrates how thermally tunable bending elasticity controls osmotically driven deformations in lipid bilayer membranes.

\subsection{Benchmarked lipid bilayer:}
The coarse-grained lipid membrane model was benchmarked against the Cooke–Deserno reference model \cite{p:Cooke2005a,p:Cooke2005b} to validate its thermodynamic behavior. Systematic variation of the interaction strength $w_{c}$ and temperature reveals three distinct bilayer phases: gel, fluid, and unstable. The resulting phase diagram is shown in Figure 1.

\begin{figure}[!h]
\includegraphics[width=\linewidth]{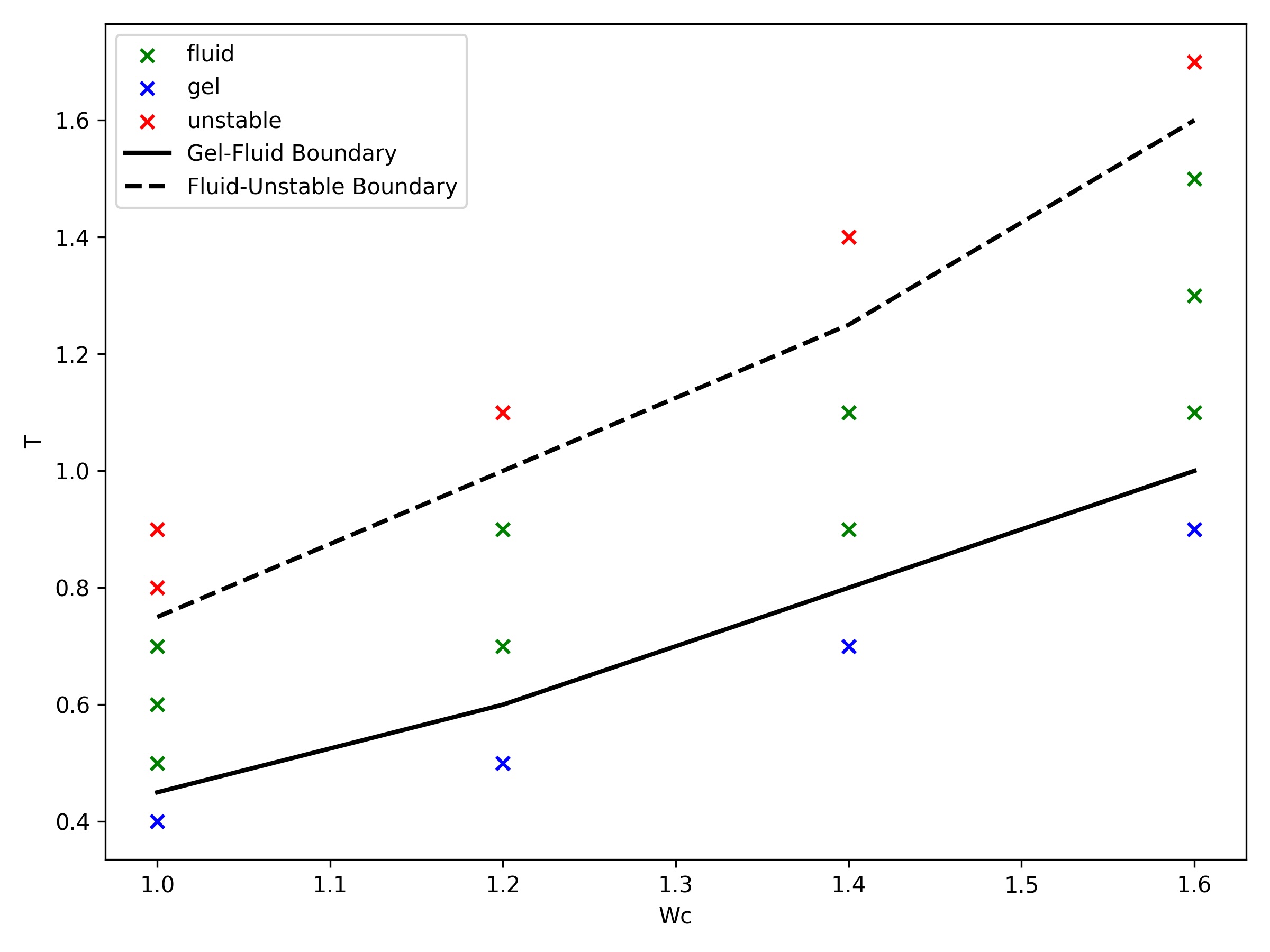}
\caption{Phase diagram of the coarse-grained membrane model benchmarking the LAMMPS simulations against the Cooke-Deserno model~\cite{p:Cooke2005a,p:Cooke2005b}.}\label{fig:phase}
\end{figure}

\subsection{Fluctuation Spectrum Analysis}

Vesicle mechanical properties were extracted from production trajectories by computing the spherical-harmonic fluctuation spectrum of the vesicle mid-surface. Head beads (type = 1) were selected and recentered each frame to remove global translation before conversion to spherical coordinates $(r,\theta,\phi)$.

The mid-surface was reconstructed using angular binning on a $N_\theta \times N_\phi$ grid ($24 \times 48$). Within each angular bin, inner and outer leaflet radii were estimated from the radial distribution using percentile statistics ($\mathrm{PIN}=15$, $\mathrm{POUT}=85$), and the mid-surface radius was defined as
\begin{equation}
r_{\mathrm{mid}}(\theta,\phi)
=
\frac{1}{2}
\left[
r_{\mathrm{in}}(\theta,\phi)
+
r_{\mathrm{out}}(\theta,\phi)
\right].
\end{equation}
The reference radius for each frame was taken as
\begin{equation}
R_0 = \langle r_{\mathrm{mid}}(\theta,\phi) \rangle_{\theta,\phi},
\end{equation}
and the dimensionless height field was defined as
\begin{equation}
u(\theta,\phi)
=
\frac{r_{\mathrm{mid}}(\theta,\phi) - R_0}{R_0}.
\end{equation}

The fluctuation field was expanded in spherical harmonics $Y_{lm}$, and the $m$-averaged spectrum was computed as
\begin{equation}
\langle |u_l|^2 \rangle
=
\frac{1}{2l+1}
\sum_{m=-l}^{l}
|a_{lm}|^2,
\end{equation}
with averaging performed over all analyzed frames.

For $l \ge 2$, the spectrum was fit to the continuum quasi-spherical membrane theory including bending and tension contributions, with an additive noise floor $C$:
\begin{equation}
\langle |u_l|^2 \rangle = \frac{k_B T}{\kappa (l-1)l(l+1)(l+2)+(\Sigma R_0^2)\,l(l+1)}+C.
\end{equation}

\begin{figure}
\includegraphics[width=\linewidth]{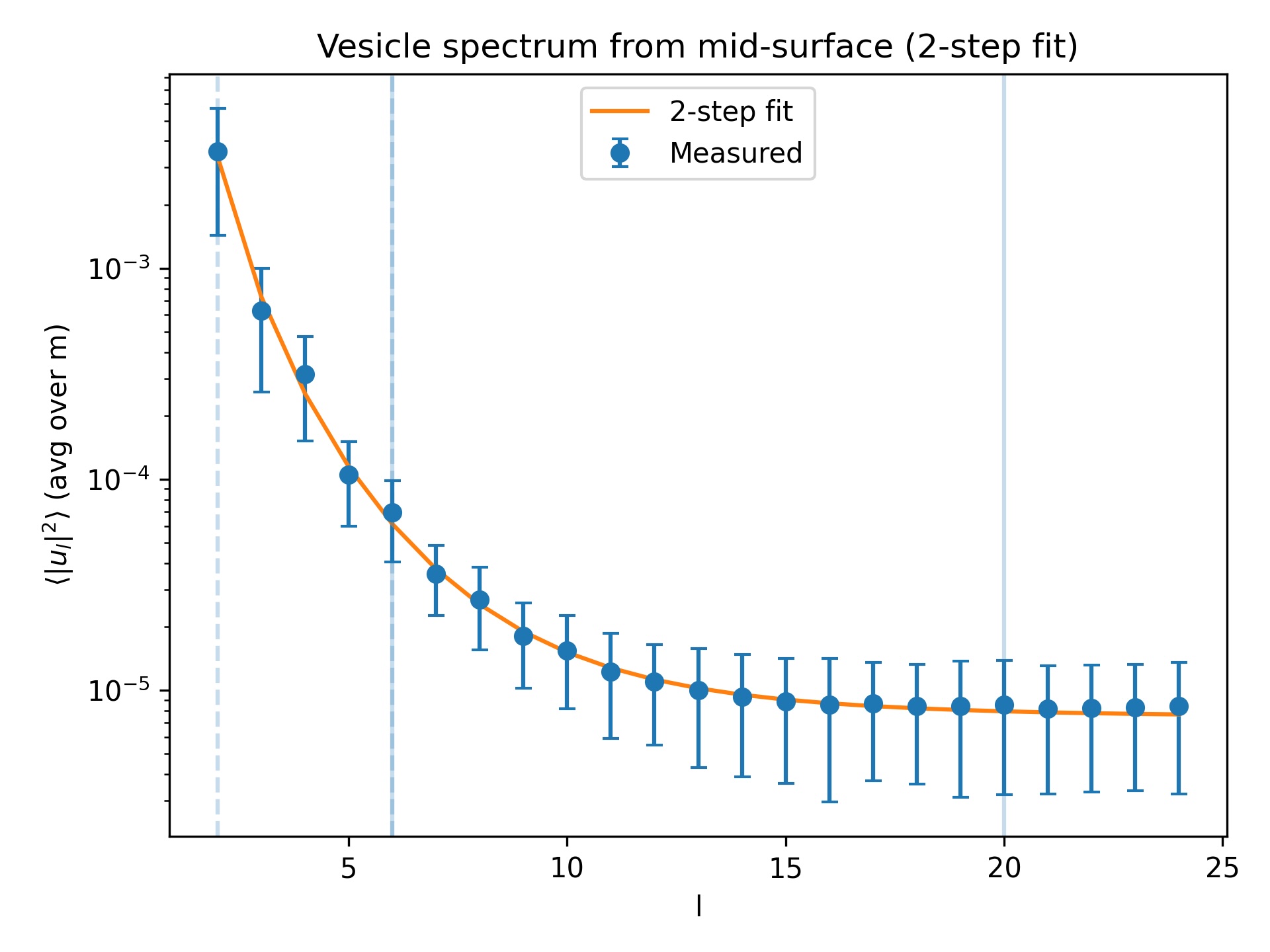}
\caption{Spherical-harmonic fluctuation spectrum of the vesicle mid-surface. Points represent the measured mode amplitudes $\langle |u_l|^2 \rangle$ averaged over production frames, with error bars indicating the standard deviation across frames. The solid line shows the two-step fit to the continuum membrane model including bending rigidity $\kappa$, surface tension $\Sigma$, and an additive noise floor $C$. Vertical lines indicate the high-$l$ and low-$l$ fitting windows used to separately determine $\kappa$ and $\Sigma$.}\label{fig:fluctuation_spectrum}
\end{figure}

To improve parameter robustness, a two-step fitting protocol was employed. First, $\kappa$ and $C$ were determined from a bending-dominated high-$l$ window ($l=6$--20) with $\Sigma R_0^2$ fixed to zero. Second, $\Sigma R_0^2$ was obtained from low-$l$ modes ($l=2$--6) while holding $\kappa$ and $C$ fixed. The surface tension was then calculated as
\begin{equation}
\Sigma = \frac{\Sigma R_0^2}{R_0^2},
\end{equation}
with $R_0$ averaged over frames. Parameter uncertainties were obtained from the covariance matrices of nonlinear least-squares fits. The resulting fluctuation spectrum and two-step fit are shown in Fig.~\ref{fig:fluctuation_spectrum}.

\subsection{Asphericity Analysis}

Vesicle shape anisotropy was quantified using the relative shape anisotropy 
$\kappa^2$ computed from equilibration and production trajectories, which were combined into a single continuous time series. Tail beads were used to construct the vesicle surface. If the production run restarted its timestep counter, timesteps were shifted to ensure continuity.

For each frame, a mid-surface point cloud was constructed using angular binning on a 
$N_\theta \times N_\phi$ grid ($30 \times 60$). Particle positions were recentered, converted to spherical coordinates, and classified into inner and outer leaflets based on the median radial distance. Within each angular bin, leaflet-resolved radii and directions were averaged and combined to define the mid-surface geometry,
\begin{equation}
r_{\mathrm{mid}} = \frac{1}{2}\left(r_{\mathrm{in}} + r_{\mathrm{out}}\right),
\qquad
\mathbf{x}_{\mathrm{mid}} = \mathbf{x}_{\mathrm{COM}} + r_{\mathrm{mid}}\,\hat{\mathbf{u}}_{\mathrm{mid}}.
\end{equation}

The inertia tensor of the mid-surface point cloud was computed about its center-of-mass, and its eigenvalues $\lambda_z \ge \lambda_y \ge \lambda_x$ were used to evaluate the relative shape anisotropy,

\begin{equation}
\quad 
\kappa^2 =
\frac{3}{2}
\frac{\lambda_x^2 + \lambda_y^2 + \lambda_z^2}
{(\lambda_x + \lambda_y + \lambda_z)^2}
-
\frac{1}{2}
\end{equation}

\begin{figure}
\includegraphics[width=\linewidth]{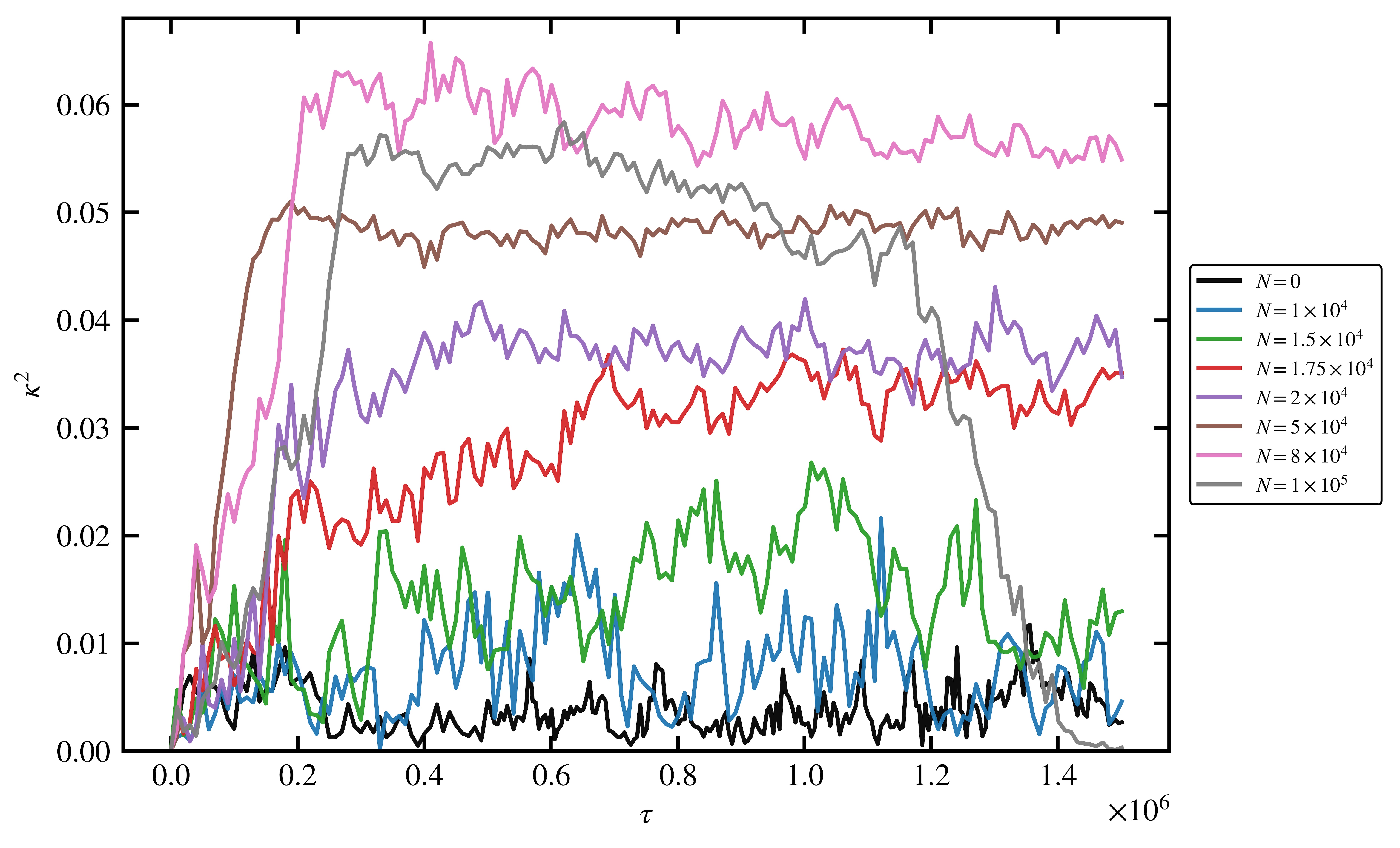}
\caption{Asphericity parameter $\kappa^2$ as a function of time for vesicle simulations with different osmolyte counts. We identify a critical osmolyte concentration $N_c \approx 1.75\times10^4$ using which we compute the critical pressure $\Delta p_c$.}\label{fig:asphericity_plot}
\end{figure}

The resulting $\kappa^2$ was reported as a function of timestep over the combined trajectory. The time evolution of the asphericity parameter $\kappa^2$ for all osmolyte concentrations considered in this work is shown in Fig.~\ref{fig:asphericity_plot}.

\subsection{Vesicle Volume Calculation}

The enclosed vesicle volume was computed directly from membrane bead configurations using a three-dimensional voxel reconstruction approach. For each frame, membrane bead coordinates were mapped into the primary simulation box and recentered to ensure the vesicle was positioned away from periodic boundaries.

A cubic grid of spacing $\Delta x$ was superimposed over the simulation box, and voxels containing membrane beads were identified as part of the membrane surface. To ensure a closed surface representation, the membrane mask was slightly expanded to eliminate small gaps arising from discretization. A flood-fill algorithm was then applied starting from the box boundaries to identify all voxels connected to the exterior region. Voxels not connected to the exterior and not belonging to the membrane were classified as interior.

The vesicle volume was estimated as
\begin{equation}
V_{\mathrm{ves}} = N_{\mathrm{inside}}\,(\Delta x)^3,
\end{equation}
where $N_{\mathrm{inside}}$ is the number of interior voxels. Reported volumes correspond to time-averaged values computed over the production window after discarding initial equilibration frames. Statistical uncertainties were estimated from the standard error of the mean over the averaging window.

\bibliographystyle{apsrev4-2}
\bibliography{arxiv}

\begin{thebibliography}{75}
\bibitem [{not()}]{note1}%
  \BibitemOpen
  \href@noop {} {}\bibinfo {note} {For unimolecular fluids, we assume $\chi >
  \chi_c = 2$. The volume of a solute molecule $v_p \approx 0.13 nm^3$ is taken
  to correspond to a glucose molecule.}\BibitemShut {Stop}%
\end{thebibliography}

\end{document}